# Aim Low, Shoot High: Evading Aimbot Detectors by Mimicking User Behavior

Tim Witschel
Technische Universität Braunschweig

Christian Wressnegger
Karlsruhe Institute of Technology

## ABSTRACT

Current schemes to detect cheating in online games often build on the assumption that the applied cheat takes actions that are drastically different from normal behavior. For instance, an Aimbot for a first-person shooter is used by an amateur player to increase his/her capabilities many times over. Attempts to evade detection would require to reduce the intended effect such that the advantage is presumably lowered into insignificance. We argue that this is not necessarily the case and demonstrate how a professional player is able to make use of an adaptive Aimbot that mimics user behavior to gradually increase performance and thus evades state-of-the-art detection mechanisms. We show this in a quantitative and qualitative evaluation with two professional "Counter-Strike: Global Offensive" players, two open-source Anti-Cheat systems, and the commercially established combination of VAC, VACnet, and Overwatch.

## 1 INTRODUCTION

In the online gaming industry vendors and operators have a natural interesting in preventing cheating. On the one hand, honest players lose interest in the game if the game objective is undermined to an extent that leaves the impression behind that there is no point in even playing. On the other hand, cheats might be used to gather in-game assets in order to, for instance, sell them in the real world [19]. Both directly affect the monetization opportunities of the game: Paying customers (the players) might leave the platform, and alternative markets that sell virtual goods are in direct competition with the vendor's in-app purchases. Moreover, e-sport tournaments have evolved to huge events that conclude sponsorship contracts, sell broadcasting rights, and distribute trophy money of millions of US Dollars [3].

Popular genres, for instance, are battle arena games such as "Dota 2" and first-person shooters like "Counter-Strike: Global Offensive" (CS:GO). According to Steam statistics, in 2019 these games had 745 k and 600 k users per day on average [2]. By these numbers, one can easily grasp the influence and market value of such games.

The type of cheat, of course, very much depends on the type of game. In first-person shooters, for instance, tools that assist in aiming and shooting at an opponent are particularly popular to enhance a player's capabilities. This includes different forms and manifestations, such as aim-locking, aim and trigger bots, and recoil reduction, which are collectively termed as "Aimbots" in common usage [12].

In the past a number of different approaches have been considered to detect cheating entities in games [e.g., 1, 6, 10, 13] and Aimbots in particular [11, 20], many of which use data mining or machine learning tools. Also Valve, the developer of the aforementioned games, has recently started to use deep learning to detect cheating users in "Counter-Strike: Global Offensive" [12]. These systems rely on detecting the difference between normal user characteristics and observed behavior. If these diverge too strongly, a cheat is detected. This is particularly striking for inexperienced players that suddenly become extraordinary effective or bots that perform actions in a flash. Current detectors thus focus on these situations and brush off more subtle scenarios where improvement is less striking. Liu et al. [11], for instance, state that Aimbots would only be able to evade detection if they *"degrade their performance significantly and play like average players"*. We however believe that this is not true and argue that adaptive Aimbots, that gradually improve aiming and shooting performance, are particularly effective for skillful players, that already perform well but use assistance to proceed to the next level. The average number of matches won in the two highest CS:GO ranks are only 5 % apart [5, 17]. Dishonest professional players are more likely to cause financial damage in a tournament setting, while amateurs using Aimbots might spoil the game for other users on a wider basis, such that both need to be equally countered.

To motivate future research on detecting Aimbots, that are used by amateurs and in particular by professional players, in this paper, we explore the feasibility of adaptive Aimbots that mimic user improvement to evade state-of-the-art detection mechanisms. To this end, we look upon first-person shooter games and define player profiles using properties that capture user behavior with respect to aiming as well as shooting capabilities. Based on these, we then define gradual improvement of the individual features with the objective of increasing the hit count. In an evaluation on a private "Counter-Strike: Global Offensive" server, we demonstrate the evasion of two open-source systems, "COW Anti-Cheat" [4] and "Sourcemod Anti-Cheat" [14], before we proceed to play on an official server secured by VAC [18], VACnet [12], and Overwatch [16]. To simulate a gain in ranks, we operate on a hit-count improvement of 5 % and show that also in the second experiment, in a total of 60 matches (about 45–60 minutes each) with two experienced players in separate teams of five, our Aimbot went unnoticed and the associated accounts remained active up until today.

Finally, we have conducted a quantitative and qualitative evaluation of the influence of the automatic adaptations on professional gameplay, showing that although the overall hit count could be improved, interventions may distract and even disrupt the flow of play.

In summary, we make the following contributions:

- **Aimbots for professional players.** We are the first to address the scenario of Aimbots used by skillful players to incrementally improve performance in a stealthy way.
- **Evasion of state-of-the-art detection.** We demonstrate the successful evasion of commercial Anti-Cheat mechanisms that are currently deployed on official game servers by mimicking gradual user improvement.
- **Open-source tools.** To foster future research and improve existing Aimbot detectors, we make all our implementations for recording player profiles and mimicking user behavior publicly available at:

    https://intellisec.org/research/aimbots

The remainder of the paper is structured as follows: In Section 2, we first describe different properties for generating user profiles and we present ways to mimic these to subvert detection in Section 3. The evaluation of the employed mechanism is presented in Section 4, demonstrating our Aimbots stealthiness in practice, before we discuss related work in Section 5. Section 6 concludes the paper.

## 2 PLAYER PROFILES

To characterize a player, we use a comprehensive set of properties describing different aspects of first-person shooter gameplay. These properties are associated with either aiming or actual shots fired, and are summarized in Table 1.

Properties $p_{a1}$ & $p_{a2}$ and $p_{s1}$ & $p_{s2}$ describe features that, according to Liu et al. [11], are particularly discriminating for human players, such as the position of player and target ($p_{a1}$), body location of the first hit ($p_{s1}$), the time to hit ($p_{a2}$), and the hit ratio ($p_{s2}$). Popular anti-cheat systems, such as "COW Anti-Cheat" [4] for instance, moreover use the initially targeted body location ($p_{s3}$) and the overall hit ratio ($p_{s4}$) as features. We extend the latter to capture the hit ratio on first attempt ($p_{s5}$) and also record the time needed for aiming at an opponent ($p_{a3}$), which in turn is closely related to property $p_{a2}$ mentioned earlier. Both are combined in property $p_{a4}$ that captures the time between defeating one and aiming at another opponent, which is a crucial skill for a professional player and is easily automated using Aimbots. To further refine these, we additionally propose properties $p_{a5}$ to $p_{a7}$ to also account for typical behavior in handling weapons when aiming.

By describing the recoil pattern Khalifa [8] has proposed a rarely considered trajectory-based feature, that however enables to very well draw conclusions about the shooter. We incorporate this as property $p_{s6}$ and also use similar trajectories during aiming ($p_{a8}$). In direct consequence, we consider the first shot fired on such a trajectory for property $p_{s7}$.

Subsequently, we define these properties in detail and describe how to record them. Section 2.1 focuses on those that concern the player's process of aiming and Section 2.2 involves those that relate to the shots actually fired.

### 2.1 Aiming

In this section, we focus on the primary task of an Aimbot: the aiming. Next to two properties from related work [11], $p_{a1}$ and $p_{a2}$, we introduce six more characteristics, $p_{a3}$ to $p_{a8}$, that are crucial elements of first-person shooter gameplay.

*2.1.1 Divergence of aiming upon coming into conflict.* A skillful player anticipates the movement of opponents and thus is able to react faster. Liu et al. [11] use this feature to identify amateurs. Together with observations of efficient eliminations, this may be an indicator of an Aimbot. Property $p_{a1}$ measures the deviation of the character's line of gaze, $\mathbf{a} \in \mathbb{R}^3$, to the viewing direction, $\mathbf{b} \in \mathbb{R}^3$, that is necessary to see the closest body part of his/her opponent,

$$\delta = \arccos\left(\frac{\mathbf{a} \cdot \mathbf{b}}{|\mathbf{a}||\mathbf{b}|}\right),$$

and we compute the average over $n$ events.

*2.1.2 Time to kill.* A rather straight-forward metric is the "effectivity of a player" in terms of defeating his/her opponent. For their detector, Liu et al. [11] measure the time between the first sighting of an opponent in the current scene, $t_1$, to a lethal shot, $t_2$. The average across $n$ events $\frac{1}{n} \sum_{i=1}^{n}(t_2 - t_1)$ denotes $p_{a2}$.

*2.1.3 Aiming duration.* For property $p_{a3}$, we break down the "time to kill" to the time actually taken for aiming at an opponent. The measurement's starting point in time, $t_1'$, is defined by the opponent entering a rather narrow field of vision, simulating the "lock region" of an Aimbot before it aims at the opponent and fires a shot. Consequently, the measurement is stopped when the player directly aims at the opponent, $t_2'$, such that $t_1 < t_1' < t_2' < t_2$ holds true. Again, the measurements are averaged over $n$ events.

*2.1.4 Duration between a kill and aiming at another opponent.* In case a scene shows multiple opponents the time between eliminating one opponent, $t_2$, and aiming at the next one, $t_3'$, can be a revealing feature of an Aimbot. An human player is not able to immediately jump between opposing viewing angles of multiple battles in short time. A stealthy Aimbot hence needs to delay operation, meaning, handing over control to the player before assisting in the next battle or attempt.

*2.1.5 Aiming with unloaded weapon.* Often the point in time the player stops pressing the fire button and stopping to aim are different. For instance, in the heat of the moment, human players might keep aiming at an opponent and attempting to shoot although there is no more ammunition left in the magazine. Many Aimbots, in turn, immediately stop tracking the target once the player ran out of bullets. To account for this, property $p_{a5}$ measures the time between the magazine becoming empty, $t_{\text{empty}}$, and the point in time the player stops aiming at an opponent, $t_4$. As for the previous measures the time duration is averaged over $n$ events of that kind.

*2.1.6 Switching between primary and secondary weapon.* Depending on the particular situation a professional player chooses between either reloading the primary weapon or switching over to another one if the magazine is empty. The latter generally is much faster and thus is chosen in case the character is not able to cover or the opponent would start firing back and causing damage. Property $p_{a6}$ defines the probability of sticking to the primary weapon and reload, rather than a fast switch to the secondary weapon, $P(X = \text{reload})$. The exact value is highly dependent on the player such that a sudden change in behavior can be revealing.

Table 1: Overview of the player profile's set of properties.

| | Id | Description | Time | Spacial | Frequency | New |
|---|---|---|---|---|---|---|
| Aiming | a1 | Divergence of aiming upon coming into conflict [11] | – | ✗ | – | – |
| | a2 | Time to kill [11] | ✗ | – | – | – |
| | a3 | Aiming duration | ✗ | – | – | ✓ |
| | a4 | Duration between a kill and aiming at another opponent | ✗ | – | – | ✓ |
| | a5 | Aiming with unloaded weapon | ✗ | – | – | ✓ |
| | a6 | Switching between primary and secondary weapon | – | – | ✗ | ✓ |
| | a7 | Time to switch to secondary weapon | ✗ | – | – | ✓ |
| | a8 | Aiming trajectory/ pattern | – | ✗ | – | ✓ |
| Shots fired | s1 | Suspiciousness of hits [11] | – | ✗ | – | – |
| | s2 | Ratio of hits when moving [11] | – | – | ✗ | – |
| | s3 | Primary body part shot at [4] | – | ✗ | – | – |
| | s4 | Hit precision [4] | – | – | ✗ | – |
| | s5 | Hit precision at first shot | – | – | ✗ | ✓ |
| | s6 | Recoil compensation [8] | – | ✗ | – | – |
| | s7 | First shot during movement | – | ✗ | – | ✓ |

*2.1.7 Time to switch to secondary weapon.* In line with the previous definition and closely related to property $p_{a5}$ ("Aiming with unloaded weapon") for $p_{a7}$ we measure the time between the magazine becoming empty, $t_{\text{empty}}$, and the switch to the secondary weapon, $t_s$, across $n$ events. In practice, this is measured simultaneously with the previous criterion. To increase the caused damage, Aimbots frequently automated the exchange of weapons such that the player can focus on the overall gameplay. A particular short time frame between emptying a magazine and switching to the secondary weapon can thus be a good indicator for Aimbots.

*2.1.8 Aiming trajectory/ pattern.* The fastest route to correct the position a player is aiming at is a direct line to the body part to hit. Many Aimbots choose exactly this naive option, which however is very different from the movement patterns seen with human players that often conduct a non-optimal path towards the opponent. Consequently, Aimbot developers have started to implement so-called spiral aim patterns to circumvent detection. Figure 1 shows both types of aiming trajectories.

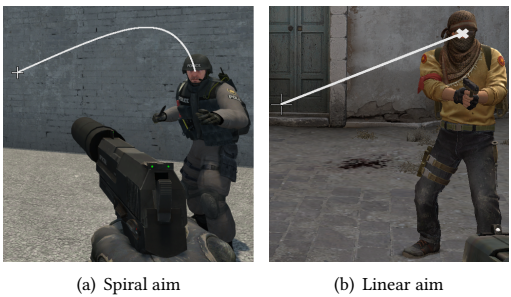

(a) Spiral aim  (b) Linear aim

Figure 1: Two different types of aiming trajectories.

For property $p_{a8}$, we thus define two measures to describe such movements. First, we capture the probability of whether the trajectory takes place above a linear movement, $P(Y = \text{above})$, and second, the height of the arch of a spiral aim as the average vertical offset during a lateral movement.

## 2.2 Shots Fired

Interestingly, the majority of related work seems to focus on properties that describe the actual shots fired during a battle rather than aiming itself: For instance, properties $p_{s1}$–$p_{s4}$ [4, 11] and $p_{s6}$ [8]. These as well as two additional properties are defined subsequently.

*2.2.1 Suspiciousness of hits.* The more critical hits succeed, the faster an opponent can be defeated. Landing such hits naturally is more difficult than hits that cause less damage. Liu et al. [11] thus model the "suspiciousness" of a critical hit as the relative position $i_c$ within a sequence of hits $s$ that led to defeating the opponent:

$$v = \begin{cases} \frac{1}{i_c} & \text{critical hit in } s \\ 0 & \text{otherwise} . \end{cases}$$

For property $p_{s1}$ the value $v$ is averaged over $n$ events/sequences.

*2.2.2 Ratio of hits when moving.* Property $p_{s2}$, that has also been proposed by Liu et al. [11], measures the probability of hitting the target when the character is moving. Many first-person shooters are designed such that aiming is easier when the character stands still or even kneels down, such that professional players do exactly this, while amateurs do not. If a player is able to hit the target reliably although he/she is moving, this usually is out of the ordinary.

*2.2.3 Primary body part shot at.* To determine the variance of the shots fired, "COW Anti-Cheat" [4] records the number of hits on individual body parts. Additionally, over $n$ events we record the probability of whether the shooter hits the body part that is closest to the initial position of the hairline cross as property $p_{s3}$. Having multiple hits on the nearest body part is extremely effective, which however is usually only achieved by an Aimbot.

*2.2.4 Hit precision.* The probably most straight-forward measure, next to property $p_{a2}$ ("Time to kill"), is the precision of the shots fired at an opponent [4]. Similarly to the previous definition, for property $p_{s4}$ we determine the number of shots on the target, but only consider the ratio of shots that hit any body part rather than missing the opponent at all to the total number of shots fired.

*2.2.5 Hit precision at first shot.* In addition to the overall hit precision, for property $p_{s5}$ we only consider the first shot in order to capture the probability of successfully hitting the opponent at the first attempt. A large value indicates a more effective shooter. However, extraordinarily high values usually are only achieved with assistance.

*2.2.6 Recoil compensation.* When firing an arm in a first-person shooter game, each weapon has a specific recoil pattern with individual characteristics, which influences the ability to aim. The longer one is firing, the larger is the recoil. Figure 2(a) shows these characteristics on a two-dimensional map for two different weapons in "Counter-Strike: Global Offensive". The first shot is marked by the hairline cross and the color gradient (orange to red) indicates the sequence of shots fired.

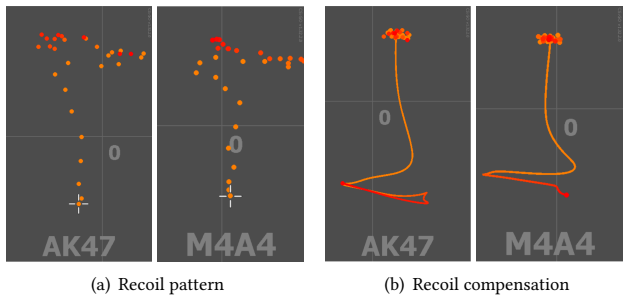

(a) Recoil pattern    (b) Recoil compensation

**Figure 2: Recoil patterns and compensation [8].**

To compensate for recoil patterns and offer a "steady hand", Aimbots automatically counteract these characteristic movements by implementing fixed compensation curves, such as shown in Figure 2(b). For property $p_{s6}$, we thus calculate the average compensation of $k$ shots fired per $n$ events as the ratio of the difference of the initial line of gaze, $\mathbf{a} \in \mathbb{R}^3$ (cf. Section 2.1.1), and the viewing angle for each shot fired, $\mathbf{a_j} \in \mathbb{R}^3$, to the location of the hit, $\mathbf{c_j} \in \mathbb{R}^3$:

$$comp = \sum_{j=1}^{k} \frac{|\mathbf{a} - \mathbf{a_j}|}{\mathbf{c_j}}$$

*2.2.7 First shot during movement.* Finally, we come back to aiming movements and consider the point in time an Aimbot starts firing on the trajectory towards an opponent. Usually, assistance of an Aimbot is started by pressing the fire button, such that the character would immediately start shooting. However, only average players start shooting right away, while skillful players wait till the body part to hit is lined up with his/her sights. Property $p_{s7}$ thus measures the average divergence of the character's line of gaze, $\mathbf{a} \in \mathbb{R}^3$, and the location of the first shot, $\mathbf{c_1} \in \mathbb{R}^3$, as the angle between these vectors, $\frac{1}{n} \sum_{i=1}^{n} \angle(\mathbf{a}, \mathbf{c_1})$.

## 3 MIMICKING USER IMPROVEMENT

Improving the performance of an online gamer can be achieved in various ways and all to often rather naive mechanisms are employed that are easily detected. For evading automatic detection, we thus propose to mimic genuine user improvement. To this end, our Aimbot first records game statistics according to the previously defined properties (Section 3.1), based on which the Aimbot then provides assistance within the capabilities of the player (Section 3.2). Subsequently, we describe these steps in more detail.

### 3.1 Recording

In a bootstrapping phase the Aimbot records data from genuine gameplay to construct a user-specific profile that is then used as basis for (gradual) improvement. In order to build a model that is as expressive as possible, the following three criteria need to be fulfilled:

**C1. Game duration.** The player is asked to carry out 12 hours of battles in a competitive setting, which correspond to about 16 matches on average. Additionally, at least this average number of games needs to be played to prevent recording data from a few, exceptionally long matches.

**C2. Number of games won.** At least 10 of these games need to be won by the player in question such that the profile is not built from bad performances. This is the same number of matches Valve asks for, before automatically assigning a skill group or rank [17].

**C3. Number of samples.** Each of the individual properties defined in Section 2 requires a minimal number of data points to reliably model the associated behavior it has been designed for. The specific values, in turn, have been determined experimentally.

Only the combination of all the above criteria ensures reliable recordings. Especially the third one has been shown to be crucial in practice: Occasionally, even during a rather long game there might not be much contact with an opponent, which renders acquiring the necessary data impossible. Similarly, if no critical hit is landed ($p_{s1}$) and no or not enough switches between primary and secondary weapon are performed ($p_{a6}$ & $p_{a7}$) these characteristics can not be sufficiently modelled.

### 3.2 Imitation

Automated user improvement needs to be executed in a way that prevents static rules or even a statistical classifier from grasping a conclusive difference to genuine gameplay. We achieve this through randomized changes to the properties of the player's profile with respect to the data recorded earlier, that however increases performance in the long run. For instance, in the case of property $p_{s3}$ ("Primary body part shot at") the Aimbot randomly chooses the body part to target such that the overall distribution of the property does not change. However, a player that constantly improves in each and every property might still draw attention to detection mechanisms. To circumvent this, we additionally randomize the growth of each characteristic: In 60 % of the cases the property is improved to ensure an overall positive progression. In 30 % of time, though, the value degrades and in 10 % there is no change at all.

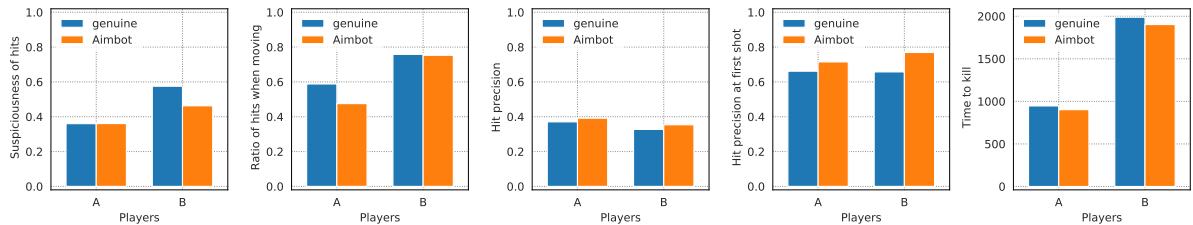

Figure 3: Improvement during adaptive adjustment of properties $p_{s1}$, $p_{s2}$, $p_{s4}$, $p_{s5}$, and $p_{a2}$.

This methodology works well for most of the described properties. Yet, for some measures interdependencies to other properties need to be considered. For instance, when adjusting the aiming trajectory to strike a balance between linear and spiral aim ($p_{a8}$) the bot additionally smoothens the motion according to the time usually spent for the overall aiming procedure ($p_{a3}$). Moreover, there exist situations where the Aimbot must not take action at all. For instance, "Counter-Strike: Global Offensive" allows to blind opponents with flash grenades. If the active player is blinded, of course, it would be easily detectable by humans as well as automated systems, if a player would still be able to aim or shoot at an opponent with precision.

## 4 EVALUATION

We have implemented the previously described strategy of adaptively improving a player's performances based on an existing game bot for "Counter-Strike: Global Offensive", called Katebot [9]. The effectiveness of our approach in practice and its evasion capabilities are evaluated by conducting the following two experiments: First, we set up a game server as it is typically used in private domains, LAN parties, or tournaments, and additionally install publicly available Anti-Cheat tools (Section 4.1). Second, we proceed to play on official servers provided by Valve that are protected against cheating using a combination of VAC, VACnet, and Overwatch (Section 4.2).

### 4.1 Private Game Server

For our first experiment, we operate a dedicated "Counter-Strike: Global Offensive" server according to specifications provided by the vendor [15]. This environment is identically to official game servers, but is entirely isolated from the outside and does not come with anti-cheat protection initially. Hence, we additionally install "COW Anti-Cheat" [4] as well as "Sourcemod Anti-Cheat" [14], two open-source systems for detecting cheats such as Aimbots. Unfortunately, the detector by Liu et al. [11], AimDetect, has not been made available to us for examination and thus could not be included in this experiment.

**Evasion.** The number of match wins of the two highest CS:GO ranks are about 5 % apart [5]. As an initial test of our Aimbot's stealthiness, we thus play with a fixed improvement objective of 5 %. Both open-source detectors implement rule-based detection and thus are designed to report any abuses immediately. Consequently, playing a few games is enough to verify that our Aimbot is not detected. Unsurprisingly, as we have actively included some of the measures also used by these detectors (e.g., $p_{s3}$ and $p_{s4}$), our Aimbot has indeed remained unnoticed.

**CPU/Memory Consumption.** Additionally, we have run preliminary test on the consumption of resources to ensure that the overall gaming experience is not negatively influenced due to the supervision implemented by our Aimbot. Short-time freezes or lags are particularly critical for first-person shooters. Our implementation merely uses 1–2 % of additionally CPU and only 170 MiB on average, such that no perceivable influence is imposed on the players.

### 4.2 Public Game Server

As final and most crucial experiment, we now deploy our Aimbot on official "Counter-Strike: Global Offensive" game servers run by the game vendor. In contrast to private servers, these additionally employ a combination of VAC, VACnet, and Overwatch.

VAC [18] is a client-side detector that is built upon signature-based detection, checking for memory modifications, and the analysis of DNS caches to detect requests to servers associated with cheat software. VACnet [12], in turn, is a rather new server-side mechanism that analyzes behavioral patterns and learns them using methods from deep learning based on very similar features as described in Section 2. According to McDonald [12], it verifiably detects 80–95 % of the cheats. The verification is provided by Overwatch [16], a community-driven regulative system, where experienced players act as reviewers for incidents reported by other users as well as automatic detection.

To limit the ethical implications of our experiment, all games have been performed during practice sessions, such that none of the contestants suffer a loss in CS:GO ranks or even financial loss.

**Quantitative results.** For this experiment, two players conduct a total of 60 matches of about 45–60 minutes each. Each contestant plays 15 games with and 15 games without Aimbot in a team of five players (team members are randomly assigned by the game server). Similarly to the previous experiment a fixed improvement objective of 5 % above the user's profile is set. While the overall improvement was clearly noticeable in the time needed to defeat an opponent, not all properties have had equally positive response.

Figure 3 and Table 2 break down the results for five of the profile's properties in detail. Property $p_{s1}$ measures how early one achieves a critical hit during a battle, which Liu et al. [11] argue to be indicative for how suspicious hits are. Our experiments shown that it is not as simple as that and adjusting the hit rate may even influence a player negatively. The performance of Player B, for instance, decays by 19.6 % in comparison to playing genuinely. For the ratio of successful hits when moving [11], in turn, Player A seems to struggle with the automatic interventions of the Aimbot, such that precision is 19.2 % worse than before. For Player A no significant change could be observed. The overall hit precision and the precision at the first shot have been successfully improved for both players by 5 % up to even 16 %, though.

Table 2: Improvement per player.

| Id | Description | Player A | Player B |
|----|-------------|----------|----------|
| s1 | Suspiciousness of hits [11] | $+0.2\,\%$ | $-19.6\,\%$ |
| s2 | Ratio of hits when moving [11] | $-19.2\,\%$ | $-0.7\,\%$ |
| s4 | Hit precision [4] | $+5.7\,\%$ | $+8.2\,\%$ |
| s5 | Hit precision at first shot | $+8.1\,\%$ | $+16.9\,\%$ |
| a2 | Time to kill [11] | $-4.7\,\%$ | $-4.3\,\%$ |

Bottom line, the Aimbot was able to decrease the time to defeat an opponent as intended by roughly 5 %. Especially for interventions in highly dynamic actions, however, skillful players seem to struggle with external influence to different extent and dependent on their specific customs. A thorough evaluation of these influences, however, is left to future work.

**Qualitative results.** After each match the (active) players have been asked to report on their experience with the Aimbot. The overall quality of gameplay has been perceived as broadly satisfactory and the Aimbot has supported actions well. A few things did stand out, though. Skillful players, for instance, frequently conduct so-called "one-taps", single shots fired off by briefly touching the fire button to limit recoil. In its current implementation the Aimbot adjusts the proposed measures for every single shot, such that carefully placed one-taps are often rendered ineffective. We suspect this to be a one of the major reasons for the negative influence on properties $p_{s1}$ and s2.

While the following is entirely anecdotally, both players have reported that they have been complimented by other players for their good performance during matches supported by the Aimbot.

**Evasion.** There exist two aspects to cheat detection on official game servers: Either an opposing player reports one as fraud or automatic detection kicks in. In both cases, the Overwatch jury manually examines video recordings of the incident. While in a prestigious tournament, a decision is made right away, for practice sessions, as in our experiments, this may take several weeks. Up until today—several months after our experiments—the used accounts have not been flagged and remain active.

## 5 RELATED WORK

A large variety of aspects on protecting (online) games against cheating have been considered in the past. Tian et al. [19] study the state of protection mechanisms of mobile games against active modifications of the game logic on the client-side. VAC [18], for instance, is designed to fend of such basic modifications, but also detecting entire program flows that are only possible due to cheats has been shown to be feasible [1]. Many bots, however, act passively by assisting the user according to game logic. Often these are easily detectable, though, based on unnatural input dynamics [6], untypically high improvement [11], repetitive actions [10, 13], or a general deviation from normality [12]. Our approach stays clear of such artifacts and thereby surpasses other bots such as Katebot [9] and Charlatano [7], which do not adapt to human behavior.

## 6 CONCLUSION

Aimbots intervene in the gameplay of first-person shooters to assist the player in aiming at opponents. Automatic detection of such cheats builds on being able to discriminate genuine from inhuman actions. We argue that this differentiation is not necessarily existent and is particularly challenging to make for professional players. By demonstrating the evasion capabilities of gradual, user-specific improvements against commercially established detection systems, we hope to point the community towards new directions for protecting online games. Interestingly, in other fields of computer security, such as malware and attack detection, stealthiness and methods for subverting detection have a long-standing history, already.

To effectively counter this problem for user-centric applications, such as online games, merely refining the features that are used for detection likely is insufficient. Instead, game objectives that require a bot to perform repetitive and striking actions need to be established. How these may look like for first-person shooter games remains an open question.